\documentstyle[amssymb,aps,12pt]{revtex}

\begin{document}
\author{Zhi-Jian Li$^1$\thanks{%
E-mail address: zjli@mail.sxu.edu.cn}, J. Q. Liang$^1$, D. H. Kobe$^2$}
\title{Larmor precession and barrier tunneling time of a neutral spinning particle}
\address{$^1$Department of Physics and Institute of Theoretical Physics, Shanxi\\
University, Taiyuan, Shanxi 030006, People's Republic of China\\
$^2$Department of Physics, University of North Texas, Denton, Texas\\
76203-5370, USA}
\maketitle

\begin{abstract}
The Larmor precession of a neutral spinning particle in a magnetic field
confined to the region of a one dimensional-rectangular barrier is
investigated for both a nonrelativistic and a relativistic incoming
particle. The spin precession serves as a clock to measure the time spent by
a quantum particle traversing a potential barrier. With the help of general
spin coherent state it is explicitly shown that the precession time is equal
to the dwell time in both the nonrelativistic and relativistic cases. We
also present a numerical estimation of the precession time showing an
apparent superluminal tunneling.

PACS number(s): 03.65.Xp, 03.65.Ta

\vspace{0.4cm}
\end{abstract}

\section{\protect\smallskip Introduction}

The time spent by a quantum particle passing through a potential barrier has
been one of the most controversial question since the founding of quantum
physics and has attracted considerable attention from both a theoretical
perspective and a experimental view\cite{1} \cite{2}. There are various
approaches dealing with tunneling time\cite{3}, but there has been no
clear-cut answer to this old question \cite{4}\cite{5}. Recently a number of
experiments\cite{6}\cite{7}\cite{8} indicating superluminal transmission of
photons has renewed interest in this subject.

In quantum mechanics, time enters as a parameter rather than an observable.
Thus, there is no direct way to calculate tunneling time. For particles with
given energy there exist at least three different notions of the traversing
time in the literature\cite{9}\cite{10}\cite{11}, i.e., the Wigner-time, the
B\"{u}ttiker-Landauer time and the Larmor-time, corresponding to different
criteria. The Wigner time accounts for how long it takes for the peak of a
wave packet to emerge from the exit face of the tunnel barrier relative to
the time that the peak of the incident wave packet arriving at the entrance
face. The calculation of the Wigner time is based on an asymptotic treatment
of tunneling as a scattering problem utilizing the method of stationary
phase to calculate the position of the peak of a wave packet. This tunneling
time is simply the derivative of the phase of the tunneling amplitude with
respect to the energy of the particle. B\"{u}ttiker and Landauer consider
the case that the height of the barrier or the amplitude of the incident
wave is modulated sinusoidally in time. They have found that if the
frequency of the modulation is very low, the tunneling particle will see the
instantaneous height of the barrier and the transmitted waves adiabatically
follow the modulation. However, as the frequency of the modulation
increases, the transmitted waves will no longer be able to follow
adiabatically the rapidly varying modulation. The B\"{u}ttiker-Landauer time
is the modulation period such that the transmitted wave begins to depart
from an adiabatic following of the modulation.

Larmor precession was first introduced long ago as a thought experiment
designed to measure the time associated with scattering events\cite{12}.
Subsequently the method was applied to measure the tunneling time of
particles penetrating barrier with a magnetic field confined to the barrier
region, causing the spin of particle to precess\cite{13}. The original
scheme \cite{13} considered only the rotation of the spin in the plane that
is perpendicular to the magnetic field. Later it was recognized that a
particle tunneling through a barrier in the magnetic field does not actually
perform a simple Larmor precession in a plane\cite{11}. The main effect of
the magnetic field is to align the spin with the field since the particle
with spin parallel to the magnetic field has lower energy and less decay
rate in barrier region than that of particle with spin antiparallel to the
magnetic field. The total angular change of the tunneling particle divided
by the Larmor precession frequency is the Larmor time\cite{11}. The
literature also invokes a dwell time\cite{14} $\tau _d$ defined as the ratio
of integrated probability density over the barrier region to the incident
flux. The dwell time measures how long the matter wave in the barrier
regardless of whether the particle is reflected or transmitted.

In the present paper we revisit the Larmor precession of a neutral spinning
particle with a general spin coherent state as a clock to measure the
tunneling time through a barrier and extend for the first time the study of
quantum tunneling to the relativistic regime. The advantage of using the
spin coherent state is that an equation of motion for the expectation value
of spin operator in the magnetic field within a barrier is obtained and
identified with the equation of spin precession. With the help of the
equation of motion we find that both nonrelativistic and relativistic
neutral spin 1/2 particles perform a simple Larmor precession in
three-dimensional space. The Lamor precession time of spin in a magnetic
field confined to the potential barrier is compared with the time in the
absence of barrier in order to show the apparent superluminal tunneling.

\section{Tunneling time for a Nonrelativistic particle}

Consider the one-dimensional rectangular potential penetration for a neutral
particle of spin $\frac 12$ with momentum $p$ and mass $m$. The Hamiltonian
is

\begin{eqnarray}
H &=&\frac{p^2}{2m}+V_0-\frac{\hbar \omega _L}2{\normalsize \sigma }_3\text{%
\qquad }|x|<d  \nonumber \\
H &=&\frac{p^2}{2m}\text{ \qquad \qquad \qquad \qquad \thinspace }|x|>d
\end{eqnarray}
where V$_0$ is the height of the barrier situated between -d and d , $\omega
_L=\frac{2\mu B}\hbar $ is the Larmor frequency and $\mu ,$ $\hbar $ denote
the magnetic moment and Planck constant, respectively. Here the
time-independent magnetic field {\bf B }is assumed in the z-direction and is
confined to the barrier. $\sigma _{1,}\sigma _{2,}\sigma _3$ are the Pauli
spin matrices. The incoming wave is in the x-direction

\begin{equation}
\psi _i=\left( 
\begin{array}{l}
u_1 \\ 
u_2
\end{array}
\right) e^{i\kappa _0x}
\end{equation}
where $\kappa _0=\sqrt{\frac{2mE}{\hbar ^2}}$ denotes the wave number and $E$
is the energy of the particle. The component $u_1$ ($u_2$) of the incoming
wavefunction corresponds to spin up (down). We assume the incoming spinor is
a normalized spin coherent state, which is the eigenstate of spin operator $%
{\bf \sigma \cdot n}$ with unit eigenvalue\cite{15}.

\begin{equation}
{\bf \sigma \cdot n}\left( 
\begin{array}{l}
u_1 \\ 
u_2
\end{array}
\right) =\left( 
\begin{array}{l}
u_1 \\ 
u_2
\end{array}
\right)
\end{equation}
where ${\bf n=}(\sin \theta \cos \varphi ,\sin \theta \sin \varphi ,\cos
\theta )$ is an arbitrary unit vector with a polar angle $\theta $ and
azimuthal angle $\varphi $. The two components of the spinor are found to be

\begin{eqnarray}
u_1 &=&\cos \frac \theta 2e^{-i\frac \varphi 2}  \nonumber \\
u_2 &=&\sin \frac \theta 2e^{i\frac \varphi 2}
\end{eqnarray}
We only consider the case of $E<V_0$ for quantum tunneling . The wave
function to the left of the barrier ($x<-d$) is

\begin{equation}
\psi _1=\left( 
\begin{array}{l}
u_1 \\ 
u_2
\end{array}
\right) e^{i\kappa _0x}+\left( 
\begin{array}{l}
A_1 \\ 
A_2
\end{array}
\right) e^{-i\kappa _0x}
\end{equation}
and the transmitted wave function to the right of the barrier ($x>d$) is

\begin{equation}
\psi _3=\left( 
\begin{array}{l}
D_1 \\ 
D_2
\end{array}
\right) e^{i\kappa _0x}
\end{equation}
In the barrier the wave function is

\begin{equation}
\psi _2=\left( 
\begin{array}{l}
B_1e^{\kappa _1x} \\ 
B_2e^{\kappa _2x}
\end{array}
\right) +\left( 
\begin{array}{l}
C_1e^{-\kappa _1x} \\ 
C_2e^{-\kappa _2x}
\end{array}
\right)
\end{equation}
where $\kappa _{1,}\kappa _2$ are given by

\begin{eqnarray}
\kappa _1 &=&\sqrt{\frac{2m}{\hbar ^2}(V_0-E-\frac{\hbar \omega _L}2)} 
\nonumber \\
\kappa _2 &=&\sqrt{\frac{2m}{\hbar ^2}(V_0-E+\frac{\hbar \omega _L}2)}
\end{eqnarray}

The coefficients A$_i$ , B$_i$ , C$_i$ and D$_i$ ($i=1,2$) in the
wavefunction are obtained from boundary conditions $\psi _1(-d)=\psi _2(-d)$
, $\psi _2(d)=\psi _3(d)$ and $\frac d{dx}\psi _1(x)|_{x=-d}=\frac d{dx}\psi
_2(x)|_{x=-d}$ , $\frac d{dx}\psi _2(x)|_{x=d}=\frac d{dx}\psi _3(x)|_{x=d}$,

\begin{eqnarray}
D_i &=&\sqrt{T_i}e^{i\phi _i}e^{-i2d\kappa _0}u_i  \nonumber \\
A_i &=&\sqrt{R_i}e^{-i\frac \pi 2}e^{i\phi _i}e^{-i2d\kappa _0}u_i  \nonumber
\\
B_i &=&\frac{i\kappa _0+\kappa _i}{2\kappa _i}e^{id\kappa _0-d\kappa _i}D_i 
\nonumber \\
\text{\quad }C_i &=&\frac{-i\kappa _0+\kappa _i}{2\kappa _i}e^{id\kappa
_0+d\kappa _i}D_i
\end{eqnarray}
where 
\begin{eqnarray}
T_i &=&\frac{4\kappa _0^2\kappa _i^2}{(\kappa _0^2+\kappa _i^2)^2\sinh
^2(2d\kappa _i)+4\kappa _0^2\kappa _i^2}  \nonumber \\
R_i &=&\frac{(\kappa _0^2+\kappa _i^2)^2\sinh ^2(2d\kappa _i)}{(\kappa
_0^2+\kappa _i^2)^2\sinh ^2(2d\kappa _i)+4\kappa _0^2\kappa _i^2}  \nonumber
\\
\phi _i &=&\arctan (\frac{\kappa _0^2-\kappa _i^2}{2\kappa _0\kappa _i}\tanh
(2d\kappa _i))
\end{eqnarray}

For our purposes we shall consider the case of infinitesimal field limit,
such that

\begin{equation}
\kappa _1\simeq \kappa -\frac{m\omega _L}{2\hbar \kappa }\text{ , \quad }%
\kappa _2\simeq \kappa +\frac{m\omega _L}{2\hbar \kappa }\text{ , \quad }%
\kappa =\sqrt{\frac{2m}{\hbar ^2}(V_0-E)}
\end{equation}
The transmission and reflection probabilities can be expanded as the power
series of the small quantity $\frac{m\omega _L}{2\hbar \kappa }$. The first
order approximation is

\begin{eqnarray}
T_1 &=&T(\kappa _1)\simeq T(\kappa )-\frac{\partial T}{\partial \kappa }%
\frac{m\omega _L}{2\hbar \kappa },\text{ \quad }T_2=T(\kappa _2)\simeq
T(\kappa )+\frac{\partial T}{\partial \kappa }\frac{m\omega _L}{2\hbar
\kappa }  \nonumber \\
R_1 &=&R(\kappa _1)\simeq R(\kappa )-\frac{\partial R}{\partial \kappa }%
\frac{m\omega _L}{2\hbar \kappa },\text{ \quad }R_2=R(\kappa _2)\simeq
R(\kappa )+\frac{\partial R}{\partial \kappa }\frac{m\omega _L}{2\hbar
\kappa }
\end{eqnarray}
and we have

\begin{equation}
\sqrt{T_1T_2}\simeq T(\kappa ),\text{ \qquad }\sqrt{R_1R_2}\simeq R(\kappa )
\end{equation}

If we denote the transmission and reflection probabilities of spin up (down)
by $T_{+}$ ($T_{-}$) and $R_{+}$ ($R_{-}$) respectively, then it can be
easily shown that

\begin{eqnarray*}
T_{+} &\equiv &|D_1|^2=T_1|u_1|^2,\text{ \quad }T_{-}\equiv
|D_2|^2=T_2|u_2|^2 \\
R_{+} &\equiv &|A_1|^2=R_1|u_1|^2,\text{ \quad }R_{-}\equiv
|A_2|^2=R_2|u_2|^2
\end{eqnarray*}

\begin{equation}
T_{+}+R_{+}=|u_1|^2,\text{ \quad }T_{-}+R_{-}=|u_2|^2  \nonumber
\end{equation}

\begin{equation}
T_{+}+R_{+}+T_{-}+R_{-}=1
\end{equation}
which indicates the conservation of probability. From the viewpoint of
scattering, the outgoing wave packet consists of both a reflected and a
transmitted wave packets, which are separated from each other. The outgoing
wave packet must be normalized to unity, since the incoming wave packet is
normalized to unity.

The expectation values of spin for the transmitted wave $\psi _t=%
{D_1 \choose D_2}%
$ in the infinitesimal field limit are

\begin{eqnarray}
\left\langle S_1\right\rangle _t &=&\frac \hbar 2\frac{T(\kappa )\sin \theta
\cos (\phi _2-\phi _1+\varphi )}{T(\kappa )-\frac{\partial T(\kappa )}{%
\partial \kappa }\frac{m\omega _l}{2\hbar \kappa }\cos \theta }  \nonumber \\
\left\langle S_2\right\rangle _t &=&\frac \hbar 2\frac{T(\kappa )\sin \theta
\sin (\phi _2-\phi _1+\varphi )}{T(\kappa )-\frac{\partial T(\kappa )}{%
\partial \kappa }\frac{m\omega _l}{2\hbar \kappa }\cos \theta }  \nonumber \\
\left\langle S_3\right\rangle _t &=&\frac \hbar 2\frac{T(\kappa )\cos \theta
-\frac{\partial T(\kappa )}{\partial \kappa }\frac{m\omega _L}{2\hbar \kappa 
}}{T(\kappa )-\frac{\partial T(\kappa )}{\partial \kappa }\frac{m\omega _l}{%
2\hbar \kappa }\cos \theta }
\end{eqnarray}
The expectation values of spin for the reflected wave $\psi _r=%
{A_1 \choose A_2}%
$ in the infinitesimal field limit are

\begin{eqnarray}
\left\langle S_1\right\rangle _r &=&\frac \hbar 2\frac{R(\kappa )\sin \theta
\cos (\phi _2-\phi _1+\varphi )}{R(\kappa )-\frac{\partial R(\kappa )}{%
\partial \kappa }\frac{m\omega _l}{2\hbar \kappa }\cos \theta }  \nonumber \\
\left\langle S_2\right\rangle _r &=&\frac \hbar 2\frac{R(\kappa )\sin \theta
\sin (\phi _2-\phi _1+\varphi )}{R(\kappa )-\frac{\partial R(\kappa )}{%
\partial \kappa }\frac{m\omega _l}{2\hbar \kappa }\cos \theta }  \nonumber \\
\left\langle S_3\right\rangle _r &=&\frac \hbar 2\frac{R(\kappa )\cos \theta
-\frac{\partial R(\kappa )}{\partial \kappa }\frac{m\omega _L}{2\hbar \kappa 
}}{R(\kappa )-\frac{\partial R(\kappa )}{\partial \kappa }\frac{m\omega _l}{%
2\hbar \kappa }\cos \theta }
\end{eqnarray}

The second terms in the expectation values of the z-component of spin for
both reflected and transmitted waves have an obvious interpretation as given
in Ref.[11] that the particle with spin parallel to the magnetic field has
lower energy and less decay rate in barrier region than that of particle
with spin antiparallel to the magnetic field. Equations (16) and (17) show
that the spin still performs a Larmor precession around the z-axis which
cannot be realized using the special spin-polarization of an incoming
particle with polar angle $\theta =\frac \pi 2$ and azimuthal angle $\varphi
=0$. To see the spin precession explicitly we may take the sum of
expectation values of spin components for the reflected and transmitted
waves with an infinitesimal magnetic field, i.e. $\left\langle
S_i\right\rangle =\frac{<\psi _t|\psi _t>\left\langle S_i\right\rangle
_t+<\psi _r|\psi _r>\left\langle S_i\right\rangle _r}{<\psi _t|\psi
_t>+<\psi _r|\psi _r>}=\left\langle \psi _t\left| S_i\right| \psi
_t\right\rangle +\left\langle \psi _r\left| S_i\right| \psi _r\right\rangle $%
. We have

\begin{eqnarray}
\left\langle S_1\right\rangle &=&\frac \hbar 2\sin \theta \cos (\phi _2-\phi
_1+\varphi )  \nonumber \\
\left\langle S_2\right\rangle &=&\frac \hbar 2\sin \theta \sin (\phi _2-\phi
_1+\varphi )  \nonumber \\
\left\langle S_3\right\rangle &=&\frac \hbar 2\cos \theta
\end{eqnarray}
which are of the same form as for the Larmor precession of a spin in a
uniform magnetic field. To see this let us consider a neutral particle in a
uniform constant magnetic field B along the z-direction in the absence of
potential barrier. The Larmor precession is obtained by solving the
Heisenberg equation

\begin{equation}
\frac d{dt}{\bf S(}t{\bf )=}\frac 1{i\hbar }[{\bf S(}t{\bf ),}H_s]
\end{equation}
with the spin Hamiltonian

\begin{equation}
H_s=-\frac 12\hbar \omega _L\sigma _3
\end{equation}
If the initial wave function is given by the spin coherent state i.e.

\begin{equation}
\psi _i=\left( 
\begin{array}{l}
u_1 \\ 
u_2
\end{array}
\right)
\end{equation}
the expectation values of the spin components at time t are

\begin{eqnarray}
\left\langle S_1(t)\right\rangle &=&\frac \hbar 2\sin \theta \cos (-\omega
_Lt+\varphi )  \nonumber \\
\left\langle S_2(t)\right\rangle &=&\frac \hbar 2\sin \theta \sin (-\omega
_Lt+\varphi )  \nonumber \\
\left\langle S_3(t)\right\rangle &=&\frac \hbar 2\cos \theta
\end{eqnarray}

Comparing Eqs.(18) and Eqs.(22) the Larmor tunneling time $\tau_{L}$ is
obviously obtained as

\begin{equation}
\tau _L=\omega _L^{-1}(\phi _1-\phi _2)
\end{equation}

Using Eq.(11) we expand $\kappa _{1,2}$ in Eq.(10) up to the first order of
the small quantity $\frac{m\omega _L}{2\hbar \kappa }$ . the Larmor
tunneling time is found to be

\begin{equation}
\tau _L=\frac{m\kappa _0}{\hbar \kappa }\frac{4d\kappa (\kappa ^2-\kappa
_0^2)+(\kappa ^2+\kappa _0^2)\sinh (4d\kappa )]}{4\kappa _0^2\kappa
^2+(\kappa ^2+\kappa _0^2)^2\sinh ^2(2d\kappa )}
\end{equation}

We assume that the incoming particle is a neutron with energy E and the
width and height of the rectangular barrier are 2d=8$\stackrel{\text{o}}{%
\text{A}}$ , V$_0$=470MeV respectively. The Larmor tunneling time as a
function of the particle energy E is shown in Fig. 1(a). The peculiar
feature, however, characteristic of Larmor tunneling time is that it
increases with the energy of incoming particle monotonically in agreement
with the observation in Ref.\cite{16}. It is interesting to compare the
Larmor tunneling time with the Larmor time of a neutron traversing a
constant magnetic field B confined in region $-d<x<d$ , but without a
barrier. With the same procedure as that for the case with a barrier we find
that the transmission probability tends to one in the small field limit. The
Larmor time of passage through the magnetic field region in the absence of a
barrier is

\begin{equation}
\tau _L^0=\frac{2md}{\hbar \kappa _0}
\end{equation}
which is just the ratio of the traveling distance $2d$ of the spinning
particle to its speed $v=\frac{\hbar \kappa _0}m$ . Using the same parameter
as in Fig. 1(a), the plot of $\tau _L^0$ as a function of energy E is shown
in Fig. 1(b). The ratio $r=\frac{\tau _L}{\tau _L^0}$ can be smaller than
one. In other word the speed of a neutron in a barrier is larger than that
in the free space \cite{17}. For the parameters chosen here, if the speed of
incoming particle is one tenth of the speed of the light in the vacuum, the
speed of particle though the barrier would be $5.7\times 10^{15}m/s$.

The dwell time $\tau _d$ is defined as the ratio of the probability ${\rm P}%
_b$ of finding a particle within the barrier to the incident probability
flux $J_i$

\begin{equation}
\tau _d=\frac{{\rm P}_b}{J_i}
\end{equation}
The incident probability flux $J_i$ is

\begin{eqnarray}
J_i &=&-\frac{i\hbar }{2m}(\psi _i^{+}\triangledown \psi _i-\psi
_i\triangledown \psi _i^{+})  \nonumber \\
&=&\frac{\hbar \kappa_0}{m}
\end{eqnarray}
and the probability for the particle to be in the barrier is

\begin{eqnarray}
{\rm P}_b &=&\int_{-d}^d\psi _m^{+}\psi _mdx  \nonumber \\
&=&\frac{\kappa _0^2}\kappa \frac{4d\kappa (\kappa ^2-\kappa _0^2)+(\kappa
^2+\kappa _0^2)\sinh (4d\kappa )]}{4\kappa _0^2\kappa ^2+(\kappa ^2+\kappa
_0^2)^2\sinh ^2(2d\kappa )}
\end{eqnarray}
The dwell time is found to coincide with the Larmor time exactly,$\tau
_d=\tau _L$ in agreement with the result of Ref. \cite{11} for the spin
polarization perpendicular to the magnetic field. The identity $\tau _d=\tau
_L$ has also been demonstrated in Ref.\cite{18} for potential barrier of
arbitrary shape.

\section{Quantum tunneling of relativistic particle}

A relativistic neutral particle of spin $\frac 12$ with mass m and magnetic
moment $\mu $, moving in an external electromagnetic field denoted by the
field strength tensor $F_{\mu \nu }$, is described by a four-component
spinor wave function $\psi $ obeying the Dirac-Pauli equation

\begin{equation}
\lbrack \gamma ^{{\it \mu }}\frac{c\hbar }i\partial _{{\it \mu }}+mc^2+\frac %
12{\large \mu \sigma }^{{\it \mu \nu }}F_{{\it \mu \nu }}]{\normalsize \psi
=0}
\end{equation}
where $c$ is the velocity of light in vacuum, $\gamma ^\mu =(\gamma ^0,{\bf %
\gamma })$ are Dirac matrices satisfying

\begin{equation}
\{\gamma ^\mu ,\gamma ^\nu \}=2g^{\mu \nu }
\end{equation}
with g$^{\mu \nu }$=diag$(1,-1,-1,-1)$, and

\begin{equation}
\sigma ^{{\it \mu \nu }}=\frac i2[\gamma ^{{\it \mu }},\gamma ^{{\it \nu }}]
\end{equation}
It can be shown that

\begin{equation}
\frac 12\sigma ^{{\it \mu \nu }}F_{{\it \mu \nu }}=i{\bf \alpha \cdot
E-\Sigma \cdot B}
\end{equation}
where {\bf E }and{\bf \ B }are the external electric and magnetic fields, $%
{\bf \alpha }=\gamma ^0{\bf \gamma }$, $\beta =\gamma ^0$. Here we make use
of the Pauli representation

\begin{equation}
\beta =\left( 
\begin{array}{ll}
1 & 0 \\ 
0 & -1
\end{array}
\right) ,\ \alpha _i{\bf =}\left( 
\begin{array}{ll}
0 & \sigma _i \\ 
\sigma _i & 0
\end{array}
\right) ,\text{ }\Sigma _i=\left( 
\begin{array}{ll}
\sigma _i & 0 \\ 
0 & \sigma _i
\end{array}
\right) \text{ \qquad }i=1,2,3
\end{equation}
and $S_i=\frac \hbar 2\Sigma _{i\text{ }}$is the spin operator.

We again consider the one-dimensional rectangular potential barrier. The
magnetic field {\bf B } assumed in the z-direction is confined within the
barrier region. The Hamiltonian is seen to be

\begin{eqnarray}
H_D &=&c\alpha _1p_x+\beta mc^2\qquad \qquad |x|>d  \nonumber \\
H_D &=&c\alpha _1p_x+\beta [(mc^2+V_0)-\frac \hbar 2\omega _L\Sigma _3]\text{%
\qquad }|x|<d
\end{eqnarray}
When E%
\mbox{$<$}%
V$_0$, the wave function satisfying the stationary Dirac-Pauli equation

\begin{equation}
H_D\psi =E\psi
\end{equation}
is

\begin{eqnarray}
\psi _1 &=&\frac 1{\sqrt{1+f_0^2}}\left( 
\begin{array}{l}
u_1 \\ 
u_2 \\ 
f_0u_2 \\ 
f_0u_1
\end{array}
\right) e^{\frac{ik_0x}\hbar }e^{-\frac{iEt}\hbar }+\left( 
\begin{array}{l}
A_1 \\ 
A_2 \\ 
-f_0A_2 \\ 
-f_0A_1
\end{array}
\right) e^{-\frac{ik_0x}\hbar }e^{-\frac{iEt}\hbar }\text{ \quad x%
\mbox{$<$}%
-d} \\
\psi _2 &=&\left( 
\begin{array}{l}
B_1e^{\frac{k_1x}\hbar } \\ 
B_2e^{\frac{k_2x}\hbar } \\ 
-if_2B_2e^{\frac{k_2x}\hbar } \\ 
-if_1B_1e^{\frac{k_1x}\hbar }
\end{array}
\right) e^{-\frac{iEt}\hbar }+\left( 
\begin{array}{l}
C_1e^{-\frac{k_1x}\hbar } \\ 
C_2e^{-\frac{k_2x}\hbar } \\ 
if_2C_2e^{-\frac{k_2x}\hbar } \\ 
if_1C_1e^{-\frac{k_1x}\hbar }
\end{array}
\right) e^{-\frac{iEt}\hbar }\text{ \qquad \qquad 
\mbox{$\vert$}%
x%
\mbox{$\vert$}%
\mbox{$<$}%
d} \\
\psi _3 &=&\left( 
\begin{array}{l}
D_1 \\ 
D_2 \\ 
f_0D_2 \\ 
f_0D_1
\end{array}
\right) e^{\frac{ik_0x}\hbar }e^{-\frac{iEt}\hbar }\text{ \qquad \qquad
\qquad \qquad \qquad \qquad \qquad \quad x%
\mbox{$>$}%
d}
\end{eqnarray}
where 
\begin{eqnarray}
f_0 &=&\frac{ck_0}{mc^2+E}  \nonumber \\
f_1 &=&\frac{ck_1}{mc^2+V_0+E+\mu B}  \nonumber \\
f_2 &=&\frac{ck_2}{mc^2+V_0+E-\mu B}
\end{eqnarray}
and 
\begin{eqnarray}
k_0 &=&\frac 1c\sqrt{E^2-(mc^2)^2}  \nonumber \\
k_1 &=&\frac 1c\sqrt{(mc^2+V_0)^2-(E+\mu B)^2}  \nonumber \\
k_2 &=&\frac 1c\sqrt{(mc^2+V_0)^2-(E-\mu B)^2}
\end{eqnarray}

The incoming wave, i.e. the first term on the right hand side of Eq.(36), is
assumed to be a normalized spin coherent state as in Eq.(4). The
coefficients A$_i,$ B$_i,$ C$_i$ and D$_i$ ($i=1,2$) in the wavefunction are
obtained from boundary conditions $\psi _1(-d)=\psi _2(-d)$ and $\psi
_2(d)=\psi _3(d)$, namely,

\begin{eqnarray}
D_i &=&\sqrt{T_i}e^{i\phi _i}e^{-\frac{i2dk_0}\hbar }u_i  \nonumber \\
A_i &=&\sqrt{R_i}e^{-i\frac \pi 2}e^{i\phi _i}e^{-\frac{i2dk_0}\hbar }u_i 
\nonumber \\
B_i &=&\frac{if_0+f_i}{2f_i}e^{\frac{idk_0-dk_i}\hbar }D_i  \nonumber \\
\text{\quad }C_i &=&\frac{-if_0+f_i}{2f_i}e^{\frac{idk_0+dk_i}\hbar }D_i
\end{eqnarray}
where 
\begin{eqnarray}
T_i &=&\frac{4f_0^2f_i^2}{(1+f_0^2)[(f_0^2+f_i^2)^2\sinh ^2(\frac{2dk_i}\hbar
)+4f_0^2f_i^2]}  \nonumber \\
R_i &=&\frac{(f_0^2+f_i^2)^2\sinh ^2(\frac{2dk_i}\hbar )}{%
(1+f_0^2)[(f_0^2+f_i^2)^2\sinh ^2(\frac{2dk_i}\hbar )+4f_0^2f_i^2]} 
\nonumber \\
\phi _i &=&\arctan (\frac{f_0^2-f_i^2}{2f_0f_i}\tanh \frac{2dk_i}\hbar )
\end{eqnarray}

To our purpose we again consider the infinitesimal field limit 
\begin{eqnarray}
k_1 &\simeq &k-\frac{E\hbar \omega _L}{2c^2k}\text{ , \quad }f_1\simeq \frac %
k\xi -\frac \hbar {2c\xi }(\frac k\xi +\frac E{ck})\omega _L  \nonumber \\
k_2 &\simeq &k+\frac{E\hbar \omega _L}{2c^2k}\text{ , \quad }f_2\simeq \frac %
k\xi +\frac \hbar {2c\xi }(\frac k\xi +\frac E{ck})\omega _L
\end{eqnarray}
where 
\begin{equation}
k=\frac 1c\sqrt{(mc^2+V_0)^2-E^2}\text{, \quad }\xi \equiv \frac 1c%
(mc^2+V_0+E)
\end{equation}
is the zero order approximation. The transmission and reflection
probabilities can be expanded as the power series of the small quantity $%
\frac{E\hbar \omega _L}{2c^2k}$. The first order approximation is

\begin{eqnarray}
T_1 &=&T(k_1)\simeq T(k)-\frac{\partial T}{\partial k}\frac{E\hbar \omega _L%
}{2c^2k},\text{ \quad }T_2=T(k_2)\simeq T(k)+\frac{\partial T}{\partial k}%
\frac{E\hbar \omega _L}{2c^2k}  \nonumber \\
R_1 &=&R(k_1)\simeq R(k)-\frac{\partial R}{\partial k}\frac{E\hbar \omega _L%
}{2c^2k},\text{ \quad }R_2=R(k_2)\simeq R(k)+\frac{\partial R}{\partial k}%
\frac{E\hbar \omega _L}{2c^2k}
\end{eqnarray}
and we have

\begin{equation}
\sqrt{T_1T_2}\simeq T(k),\text{ \quad }\sqrt{R_1R_2}\simeq R(k)  \nonumber
\end{equation}

\begin{equation}
(1+f_0^2)(T(k)+R(k))=1
\end{equation}
The transition and reflection probabilities of spin up and down are given
respectively by

\begin{eqnarray}
T_{+} &\equiv &(1+f_0^2)|D_1|^2=(1+f_0^2)T_1|u_1|^2  \nonumber \\
T_{-} &\equiv &(1+f_0^2)|D_2|^2=(1+f_0^2)T_2|u_2|^2  \nonumber \\
R_{+} &\equiv &(1+f_0^2)|A_1|^2=(1+f_0^2)R_1|u_1|^2  \nonumber \\
R_{-} &\equiv &(1+f_0^2)|A_2|^2=(1+f_0^2)R_2|u_2|^2
\end{eqnarray}
We have probability conservation so

\begin{equation}
T_{+}+R_{+}=|u_1|^2,\text{ \quad }T_{-}+R_{-}=|u_2|^2  \nonumber
\end{equation}

\begin{equation}
T_{+}+R_{+}+T_{-}+R_{-}=1
\end{equation}

The expectation values of spin for the transmitted wave are obtained in the
infinitesimal field limit as

\begin{eqnarray}
\left\langle S_1\right\rangle _t &=&\frac \hbar 2\frac{T(k)\sin \theta \cos
(\phi _2-\phi _1+\varphi )}{T(k)-\frac{\partial T(k)}{\partial k}\frac{%
E\hbar \omega _L}{2c^2k}\cos \theta }  \nonumber \\
\left\langle S_2\right\rangle _t &=&\frac \hbar 2\frac{(1-f_0^2)T(k)\sin
\theta \sin (\phi _2-\phi _1+\varphi )}{(1+f_0^2)\left( T(k)-\frac{\partial
T(k)}{\partial k}\frac{E\hbar \omega _L}{2c^2k}\cos \theta \right) } 
\nonumber \\
\left\langle S_3\right\rangle _t &=&\frac \hbar 2\frac{(1-f_0^2)\left(
T(k)\cos \theta -\frac{\partial T(k)}{\partial k}\frac{E\hbar \omega _L}{%
2c^2k}\right) }{(1+f_{^20})\left( T(k)-\frac{\partial T(k)}{\partial k}\frac{%
E\hbar \omega _L}{2c^2k}\cos \theta \right) }
\end{eqnarray}
The reflected part reads

\begin{eqnarray}
\left\langle S_1\right\rangle _r &=&\frac \hbar 2\frac{R(k)\sin \theta \cos
(\phi _2-\phi _1+\varphi )}{R(k)-\frac{\partial R(k)}{\partial k}\frac{%
E\hbar \omega _L}{2c^2k}\cos \theta }  \nonumber \\
\left\langle S_2\right\rangle _r &=&\frac \hbar 2\frac{(1-f_0^2)R(k)\sin
\theta \sin (\phi _2-\phi _1+\varphi )}{(1+f_0^2)\left( R(k)-\frac{\partial
R(k)}{\partial k}\frac{E\hbar \omega _L}{2c^2k}\cos \theta \right) } 
\nonumber \\
\left\langle S_3\right\rangle _r &=&\frac \hbar 2\frac{(1-f_0^2)\left(
R(k)\cos \theta -\frac{\partial R}{\partial k}\frac{E\hbar \omega _L}{2c^2k}%
\right) }{(1+f_0^2)\left( R(k)-\frac{\partial R(k)}{\partial k}\frac{E\hbar
\omega _L}{2c^2k}\cos \theta \right) }
\end{eqnarray}
The sum of expectation values of spin components for the reflected and
transmitted waves with an infinitesimal magnetic field is

\begin{eqnarray}
\left\langle S_1\right\rangle &=&\frac \hbar 2\sin \theta \cos (\phi _2-\phi
_1+\varphi )  \nonumber \\
\left\langle S_2\right\rangle &=&\frac \hbar 2\frac{1-f_0^2}{1+f_0^2}\sin
\theta \sin (\phi _2-\phi _1+\varphi )  \nonumber \\
\left\langle S_3\right\rangle &=&\frac \hbar 2\frac{1-f_0^2}{1+f_0^2}\cos
\theta
\end{eqnarray}
which are formally the same as the Larmor precession equation of spin ${\bf {%
\Sigma}}$ in a magnetic field. To see this we solve the Heisenberg equation
(19) with the Hamiltonian

\begin{equation}
H_s=-\frac 12\hbar \omega _L\beta \Sigma _3
\end{equation}
and the initial wave function

\begin{equation}
\psi _i=\frac 1{\sqrt{1+f_0^2}}\left( 
\begin{array}{l}
u_1 \\ 
u_2 \\ 
f_0u_2 \\ 
f_0u_1
\end{array}
\right)
\end{equation}
The expectation values of the spin components at time t is

\begin{eqnarray}
\left\langle S_1(t)\right\rangle &=&\frac \hbar 2\sin \theta \cos (-\omega
_Lt+\varphi )  \nonumber \\
\left\langle S_2(t)\right\rangle &=&\frac \hbar 2\frac{1-f_0^2}{1+f_0^2}\sin
\theta \sin (-\omega _Lt+\varphi )  \nonumber \\
\left\langle S_3(t)\right\rangle &=&\frac \hbar 2\frac{1-f_0^2}{1+f_0^2}\cos
\theta
\end{eqnarray}
Using the approximation Eq.(43) for $k_{1,}$ $k_2$ in Eq.(39) and Eq.(42),
the Larmor tunneling time which is defined by $\tau _L=\frac{\phi _1-\phi _2%
}{\omega _L}$ is obtained as

\begin{equation}
\tau _L=\frac{f_0}{c^2k}\frac{4dk\xi E(k^2-f_0^2\xi ^2)+\hbar (ck^2+E\xi
)(k^2+f_0^2\xi ^2)\sinh (\frac{4dk}\hbar )}{4f_0^2\xi ^2k^2+(k^2+f_0^2\xi
^2)^2\sinh ^2(\frac{2dk}\hbar )}
\end{equation}

For a relativistic neutron and a rectangular potential barrier of width 2d=8$%
\stackrel{\circ }{\text{A}}$ and height V$_0$=6000MeV, the Larmor tunneling
time as a function of kinetic energy $E_k$, which in relativistic case is
defined as the total energy $E$ minus the static energy $mc^2$, is shown in
Fig. 2(a) which is similar to the nonrelativistic case (Fig. 1(a)) except
the time scale. The Larmor time through the magnetic field region in the
absence of a barrier is

\begin{equation}
\tau _L^0=\frac{2dE}{c^2k_0}
\end{equation}
which also is exactly the ratio of the traveling distance $2d$ to speed $v=c%
\sqrt{1-(\frac{mc^2}E)^2}$ . Using the same parameters as Fig. 2(a), $\tau
_L^0$ the function of the particle kinetic energy $E_k$ is plotted in Fig.
2(b). For the parameters chosen here, if the speed of incoming particle is $%
2.9\times 10^8m/s$, the speed of particle tunneling through the barrier
would be $6.4\times 10^{15}m/s$. It is interesting to see the difference
numerically between the non-relativistic expression (24) and the
relativistic formula (61). To this end, we plot the results from both
expressions in Fig.(3) with the same kinetic energy of the incoming
particles.

In the Dirac theory, the incident probability flux $J_i$ is

\begin{equation}
J_i=\psi _i^{+}c\alpha _1\psi _i=\frac{2cf_0}{1+f_0^2}
\end{equation}
and the probability for the particle to be in the barrier is

\begin{eqnarray}
{\rm P}_b &=&\int_{-d}^d\psi _m^{+}\psi _mdx  \nonumber \\
&=&\frac{f_0^2[-4dk(k^2-\xi ^2)(k^2-f_0^2\xi ^2)+\hbar (k^2+\xi
^2)(k^2+f_0^2\xi ^2)\sinh (\frac{4dk}h)]}{(1+f_0^2)k[4f_0^2k^2\xi
^2+(k^2+f_0^2\xi ^2)^2\sinh ^2(\frac{2dk}\hbar )]}
\end{eqnarray}
The dwell time $\tau _d$ is the ratio of Eq. (60) to Eq.(59),

\begin{equation}
\tau _d=\frac{f_0}{2ck}\frac{-4dk(k^2-\xi ^2)(k^2-f_0^2\xi ^2)+\hbar
(k^2+\xi ^2)(k^2+f_0^2\xi ^2)\sinh (\frac{4dk}h)}{4f_0^2k^2\xi
^2+(k^2+f_0^2\xi ^2)^2\sinh ^2(\frac{2dk}\hbar )}
\end{equation}
Using the relation 
\begin{equation}
E=\frac{c\xi ^2-ck^2}{2\xi }
\end{equation}
it is obvious that the dwell time Eq.(61) equals exactly the Larmor time
Eq.(57).

\section{Conclusion}

Using spin coherent state of an incoming particle we show that a neutral
spinning particle penetrating the potential barrier with a constant magnetic
field gives rise to a Larmor precession from which the barrier interaction
time i.e. a time length for particle to remain in the barrier is determined.
The Larmor time coincides with the dwell time in both the nonrelativistic
and relativistic cases. The numerical calculation shows that the Larmor
tunneling time can be much smaller than the time that the particle
penetrates a constant magnetic field without a barrier, which implies the
apparent superluminal tunneling.


\begin{references}
\bibitem{1}  J. T. Cushing, Found. Phys. Lett. {\bf 10,} 269 (1995).

\bibitem{2}  A. Ranfagni, P. Mugnai, P. Fabeni and G. P. Pazzi, Appl. Phys.
Lett. {\bf 58,} 774 (1990).

\bibitem{3}  E. H. Hauge and J. A. Stovneng, Rev. Mod. Phys. {\bf 61}, 917
(1989).

\bibitem{4}  R. Landauer and T. Martin, Rev. Mod. Phys. {\bf 66,} 217 (1994).

\bibitem{5}  A. M. Steinberg, Phys. Rev. Lett. {\bf 74,} 2405 (1995).

\bibitem{6}  L. J. Wang, A. Kuzmich and A. Dogariu, Nature {\bf 406,} 277
(2000).

\bibitem{7}  D. Mugnai, A. Ranfagni and R. Ruggeri, Phys. Rev. Lett. {\bf 84,%
} 4830 (2000).

\bibitem{8}  A. M. Steinberg and R. Y. Chiao, Phys. Rev. A {\bf 49,} 3283
(1994).

\bibitem{9}  E. P. Wigner, Phys. Rev. {\bf 98,} 145{\bf \ }(1955).

\bibitem{10}  M. B\"{u}ttiker and R. Landauer, Phys. Rev. Lett {\bf 49,}
1739 (1982).

\bibitem{11}  M. B\"{u}ttiker, Phys. Rev. B {\bf 27,} 6178 (1983).

\bibitem{12}  A. I. Baz, Sov. J. Nucl. Phys. {\bf 4,} 182 (1967); {\bf 5,}
161 (1967).

\bibitem{13}  V. F. Rybachenko, Sov. J. Nucl. Phys. {\bf 5,} 635 (1967).

\bibitem{14}  F. T. Smith, Phys. Rev. {\bf 118}, 349 (1960).

\bibitem{15}  J. Q. Liang and X. X. Ding, Acta Physica Sinica {\bf 8,} 409
(1999) .

\bibitem{16}  C. Bracher, M. Kleber and M. Riza, Phys. Rev. A {\bf 60}, 1864
(1999).

\bibitem{17}  T. E. Hartman, J. Appl. Phys. {\bf 33}, 3427 (1962).

\bibitem{18}  C. Bracher, J. Phys. B {\bf 30}, 2717 (1997).

{\bf Figure Caption:}

Fig.1(a) Larmor time calculated from the non-relativistic expression Eq.(24)
in a barrier as a function of the particle energy with 2d = 8$\stackrel{%
\text{o}}{\text{A}}$, m =1.67$\times $10$^{-27}$kg and V$_0$ = 470MeV.

Fig.1(b) Larmor time of Eq.(25) without a barrier as a function of the
particle energy with 2d=8$\stackrel{\text{o}}{\text{A}}$ and m = 1.67$\times
10^{-27}$kg.

Fig.2(a) Larmor time for the relativistic case Eq.(57) in a barrier as a
function of the particle kinetic energy with 2d = 8$\stackrel{\text{o}}{%
\text{A}}$, m = 1.67$\times $10$^{-27}$kg and V$_0$ = 6000MeV.

Fig.2(b) Larmor time of Eq.(58) without a barrier as a function of the
particle kinetic energy with 2d=8$\stackrel{\text{o}}{\text{A}}$ and m = 1.67%
$\times $10$^{-27}$kg.

Fig.3 Larmor times as a function of the kinetic energy for both the
non-relativistic (dotted line) and relativistic (solid line) particles with
the same barrier width 2d=8$\stackrel{\text{o}}{\text{A}}$ but various
heights: (a) V$_0$=1000MeV, (b) V$_0$=3000MeV, (c) V$_0$=8000MeV.
\end{references}
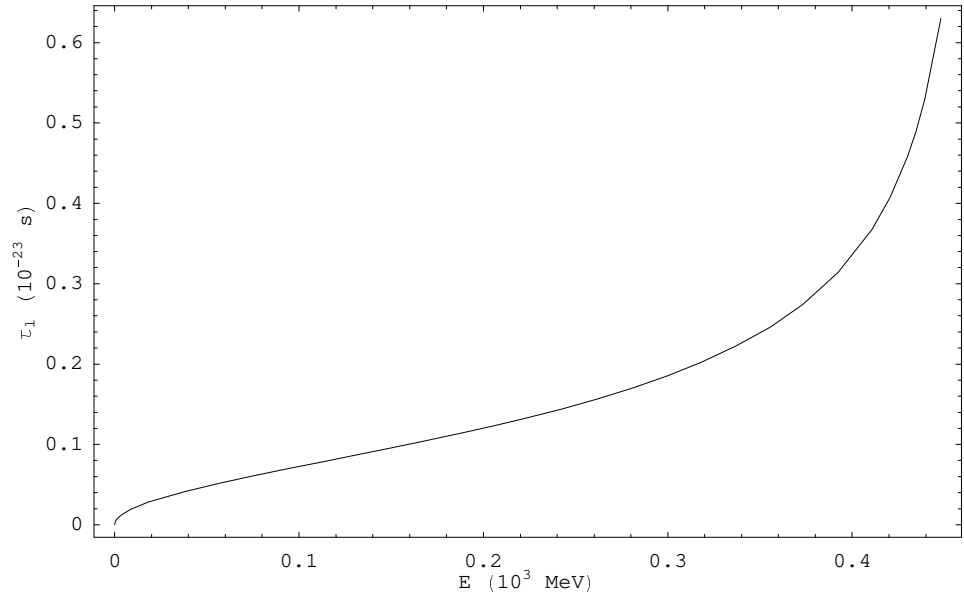
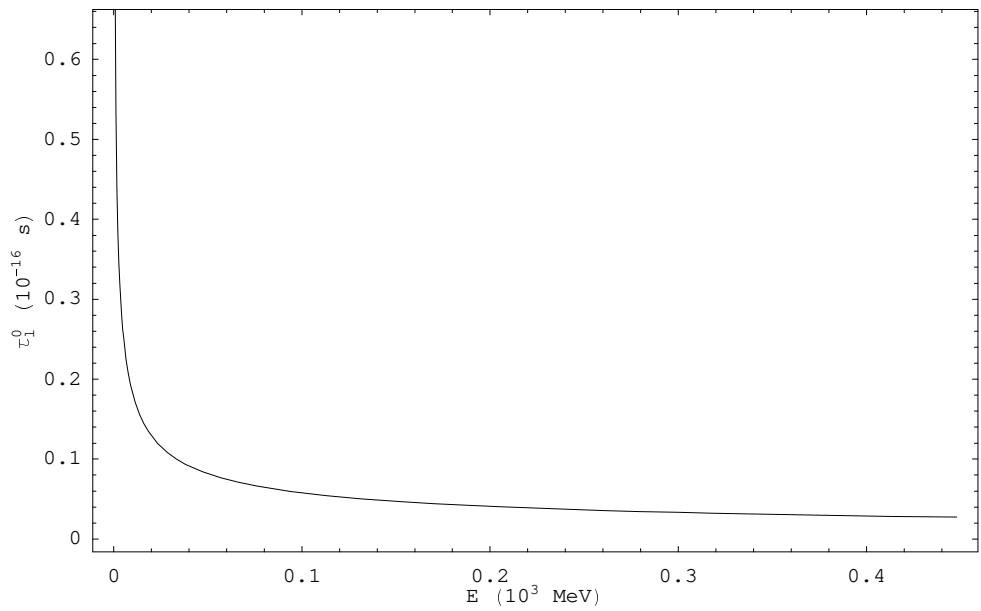
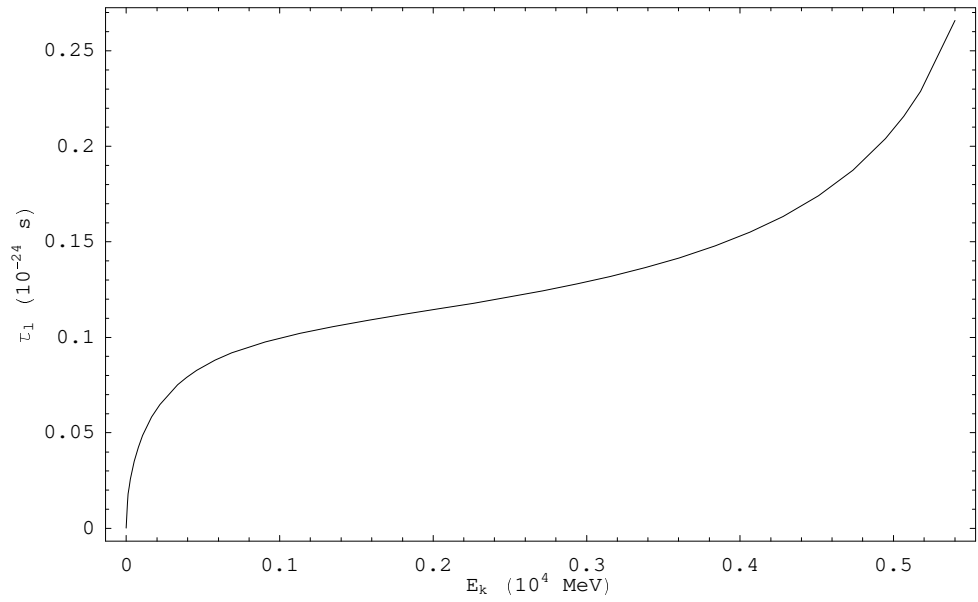
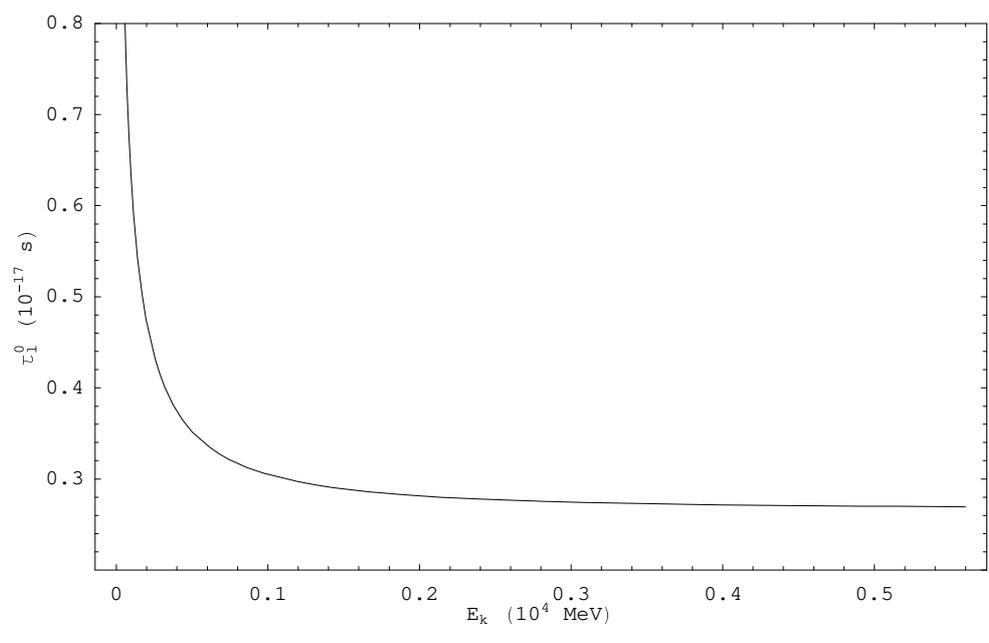
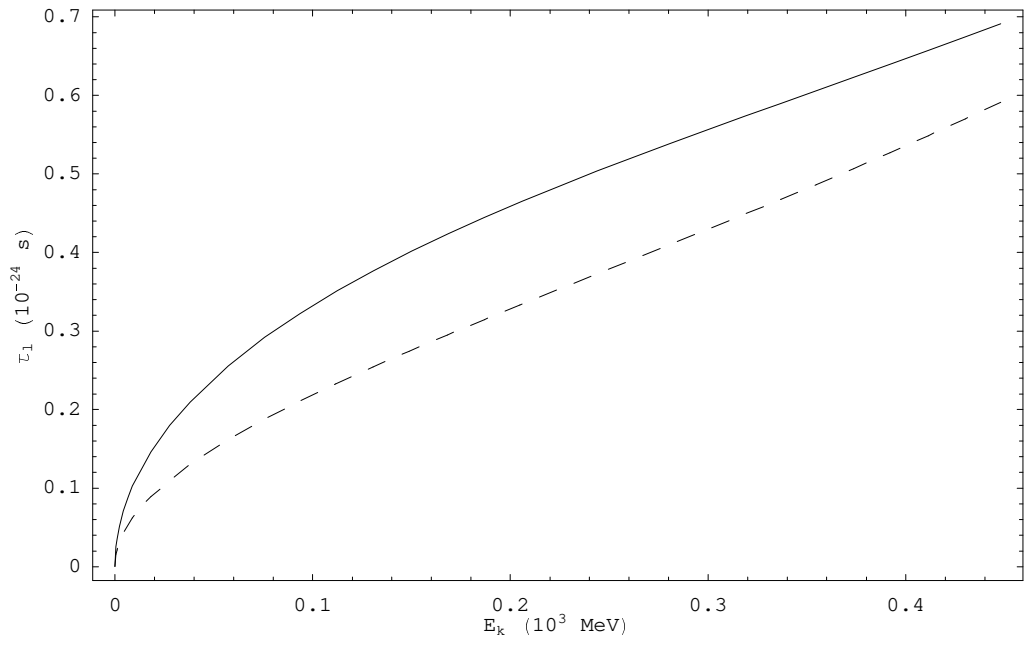
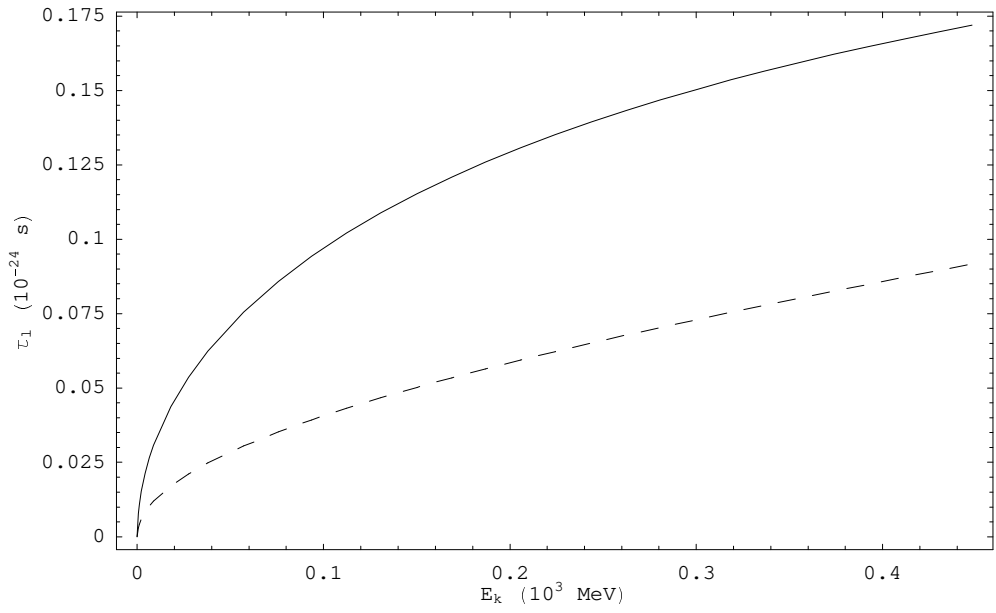
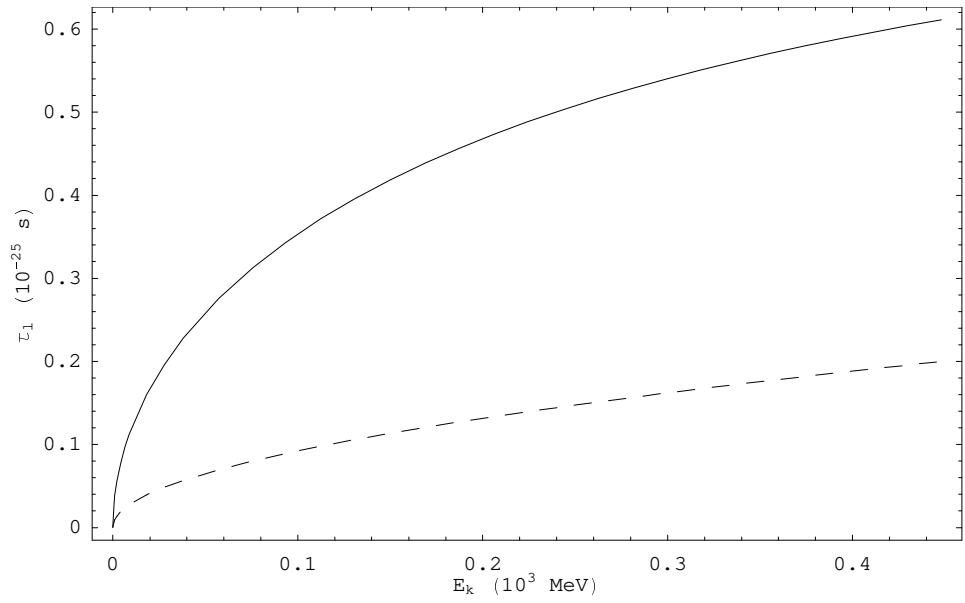
\end{document}